\documentclass[12pt,preprint]{aastex6}

\usepackage{natbib}
\usepackage{amsmath}
\usepackage{graphicx}
\usepackage{rotating}
\usepackage{url}

\shorttitle{Stability Limit of CBPs}
\shortauthors{Quarles et al.}

\begin{document}

\title{Stability Limits of Circumbinary Planets: Is There a Pile-up in the Kepler CBPs?}

\author{B. Quarles}
\affil{HL Dodge Department of Physics \& Astronomy, University of Oklahoma, Norman, OK 73019, USA}
\email{billylquarles@gmail.com}

\author{S. Satyal}
\affil{The Department of Physics, University of Texas at Arlington, Arlington, TX, 76019, USA}

\author{V. Kostov}
\affil{NASA Goddard Space Flight Center, Mail Code 665, Greenbelt, MD 20771, USA}

\author{N. Kaib}
\affil{HL Dodge Department of Physics \& Astronomy, University of Oklahoma, Norman, OK 73019, USA}

\author{N. Haghighipour}
\affil{Institute for Astronomy, University of Hawaii-Manoa, Honolulu, HI 96822, USA}

\begin{abstract}
The stability limit for circumbinary planets (CBPs) is not well defined and can depend on initial parameters defining either the planetary orbit or the inner binary orbit.  We expand on the work of Holman \& Wiegert (1999, AJ 117, 621) to develop numerical tools for quick, easy, and accurate determination of the stability limit.  The results of our simulations, as well as our numerical tools, are available to the community through \texttt{Zenodo} and \texttt{GitHub}, respectively.  We employ a grid interpolation method based on $\sim$150 million full N-body simulations of initially circular, coplanar systems and compare to the 9 known Kepler CBP systems.  Using a formalism from planet packing studies, we find that {55\%} of the Kepler CBP systems allow for an additional equal-mass planet to potentially exist on an interior orbit relative to the observed planet.  Therefore, we do \textit{not} find strong evidence for a pile-up in the Kepler CBP systems and more detections are needed to adequately characterize the formation mechanisms for the CBP population. Observations from the Transiting Exoplanet Survey Satellite are expected to substantially increase the number of detections using the unique geometry of CBP systems, where multiple transits can occur during a single conjunction. 
\end{abstract}

\section{Introduction}
The Kepler circumbinary planets (CBPs) present a rich set of dynamical systems in close binary systems that resemble architectures around single stars.  Soon after the first detection of Kepler-16b by \cite{Doyle2011}, theorists have been seeking to understand more fully the possible dynamics, evolution, and formation of these bodies \citep[e.g.,][]{Quarles2012,Meschiari2012,Kane2013,Rafikov2013,Dunhill2013}.  Analysis of the Kepler data has uncovered more CBPs around a variety of stellar hosts, such as Kepler-34b and Kepler-35b \citep{Welsh2012} that orbit nearly sunlike stars or two confirmed planets in the same system, Kepler-47b \& Kepler-47c \citep{Orosz2012b}, whose stellar hosts have a nearly circular orbit.  The Transiting Exoplanet Survey Satellite (TESS) is expected to observe $\sim$500,000 eclipsing binaries and allow for a substantial increase in the number of observed CBPs in the next few years {using the detection method outlined in \cite{Kostov2016}}.

The stability limits{, or smallest stable semimajor axis ratio,} of planets as test particles in binary systems have been identified \citep{Dvorak1986,Dvorak1989,Holman1999} assuming that the test bodies begin on nearly coplanar, circular orbits around their host stars.  However, the eccentricities of the known CBPs cover a wide range and explore regions of parameter space where the stability formula by \cite{Holman1999} (hereafter HW99) may become inadequate.  Additionally the definition of a stability limit must be inherently ``fuzzy'' due to the overlap of mean motion resonances \citep{Chirikov1979,Wisdom1980,Mudryk2006,Deck2013} and some regions of parameter space may be stable but unaccessible through processes within modern formation models.  Some studies \citep{Doolin2011,Quarles2016,Li2016} have investigated how the stability changes when the planets are significantly inclined relative to the binary planet, or at least enough to prevent the CBP from transiting \citep{Li2016}.

The evolution of these systems have been studied prior to the discovery of the Kepler CBPs using N-body dynamical models dominated by planetesimals composed of a mixture of rock and ice \citep{Quintana2006} or hydrodynamical models dominated by gas with planets embedded \citep{Artymowicz1994,Artymowicz1996,Gunther2002,ValBorro2006,Pierens2007,Pierens2008,Marzari2009}.  The actual systems do not resemble the Earth in composition, where they are typically between a $\sim$Neptune -- Jupiter mass and have volatile-rich compositions.  As such, hydrodynamical models have been used to characterize the Kepler CBPs that incorporate interactions between the growing planetary core with a gaseous disk \citep{Paardekooper2012,Meschiari2014,Kley2014,Kley2015,Bromley2015}.

When comparing the Kepler CBPs with their respective stability limits (e.g., HW99), the CBP community has remarked on the closeness of these bodies to this ``inner boundary'' and there is only
a small probability that the pile up of planets near the
stability limit is due to selection bias \citep{Li2016}.  {Although the terminology is similar, the closeness to the inner stability boundary should not be conflated with the observed 3-day pile up in the Hot Jupiters from RV observations as the underlying physical mechanisms are likely different.  This closeness in CBP systems can be quantitatively defined by: (1) a ratio of semimajor axes ($a_p/a_{c,HW}$) relative to the stability limit by HW99, (2) a spike in the distribution of planetary semimajor axes (log scale), or (3) the dynamical fullness of each system, where a dynamically full system will not allow for additional planets to be placed between the observed planet and the inner stellar binary.  In this paper, we use the third definition because the first does not account for spacing with respect to Hill spheres and the second is not currently applicable given the small number of known CBP systems.  As a result, we seek to better understand the transition to stability for CBPs, improve the historical formalism for stability, and address whether the Kepler CBPs could host additional planets on interior orbits.  In order for such planets to exist, they must presently be on mutually inclined orbits as to not transit and matching this observational constraint is beyond the scope of this work. }

Our methods, the initial conditions for our simulations, definitions for our stability analysis, and assumptions are summarized in Section \ref{sec:methods}.  In Section \ref{results}, we present the results of our numerical simulations, a comparison with the Kepler CBPs (at a population and individual level), and a discussion of how this work can be applied to future observations with TESS.  We provide the conclusions of our work and compare our results with previous studies in Section \ref{sec:conc}.

\section{Methodology} \label{sec:methods}
\subsection{Numerical Setup}
Our simulations use a modified scheme within the popular \texttt{mercury} integration package \citep{Chambers2002} that has been designed for the efficient simulation of circumbinary systems.  This modification allows for the integration of the inner binary orbit and an outer planet at different timescales while preserving the symplectic nature of the integration method.  As a result, we find that the largest integration step for the planet to be $\sim$2.5\% of the planetary Keplerian period.  Our numerical scheme stops the simulation when an instability event occurs, which we define as an intersection with the binary orbit or when the radial distance of the planet to the more massive star exceeds 10 AU.  {Using 10 AU from the more massive star as a distance cutoff is justified because the planets begin with small semimajor axes, which rules out such a large apastron distance, and planets that reach this distance are likely to exceed the respective escape velocity.}

A majority of our simulations use ideal initial conditions for the binary orbit, where the binary semimajor axis is 1 AU and the total mass $(M_A+M_B)$ of the stellar components is 1 M$_\odot$.  Our runs consider a range of binary mass ratios ($\mu = M_B/(M_A+M_B)$) from 0.01 -- 0.5 in steps of 0.01 and include one additional case, $\mu = 0.001$, for a total of 51 steps.  The eccentricity of the binary orbit varies from 0.0 -- 0.80 in steps of 0.01.  Most of our integrations begin the binary orbit at periastron ($\lambda_{bin} = 0^\circ$) because previous investigations (HW99) have shown this assumption to produce a conservative estimate for the stability limit.  However, we do investigate a small subset of runs to quantify how beginning the binary at apastron ($\lambda=180^\circ$) would change our results.

\subsection{Coplanar, Circular Planetary Orbits} \label{sec:stab}
In order to compare with previous results (i.e., HW99), we perform integrations to determine a critical semimajor axis $a_c$ for a Jupiter-mass planet that begins on an initially coplanar, circular orbit.  {We define the stability limit as the the critical semimajor axis ratio $a_c$ in units of the binary semimajor axis $a_{bin}$ and measure  it by the smallest planetary semimajor axis where a planet is stable for all choices of initial Mean anomaly or phases (i.e., the lower critical orbit, \cite{Dvorak1989}) relative to the host binary orbit.}  This definition is motivated by our models of gas giant CBPs that employ migration from a larger distance through interactions with the disk \citep{Pierens2008,Paardekooper2012,Meschiari2014,Kley2014,Kley2015,Bromley2015}.  Such studies have shown that gas disk migration around tight binaries can occur in a similar manner as in single star systems, where gas drag acts to circularize planetary orbits.  After the gas disk dissipates, the binary excites the eccentricities of close-in exoplanets leading to scattering events or expulsion from the system \citep[e.g.,][]{Silsbee2015,Kley2015,Thebault2015,Vartanyan2016}.

In order to determine $a_c$ consistently, given the above definition, we perform simulations over a grid of orbital parameters, where the total simulation time per integration is 10$^5$ binary orbits.  Our grid of planetary orbital parameters, for each combination of the stellar $\mu$ and $e_{bin}$, vary the semimajor axis ratio ($a_p/a_{bin}$) from 1.01 to 5.0 in steps of 0.01 and the initial planetary Mean anomaly from $0^\circ - 180^\circ$ in steps of $2^\circ$. 

HW99 performed their simulations using a similar definition for the stability limit but only include 8 initial phases ($0^\circ - 315^\circ$) for each test particle.  Our simulations take advantage of a symmetry with respect to the initial phase and increase the number of trial phases because some initial values can become unstable in between the $45^\circ$ increments employed by HW99.  A higher resolution is necessary to ensure that our definition for the stability limit is reliable.

\subsection{Stability Limits using the Hill Radius}\label{sec:Hill}
For very small values of both $\mu$ and $e_{bin}$, our simulations approach conditions consistent with Hill stability \citep{Szebehely1981,Gladman1993}.  This type of analysis identifies the gravitational radius of influence that the secondary mass has on the tertiary mass through the Hill radius $R_H = a(\mu/3)^{1/3}$. 

Parameterization using the Hill radius has also been used in the stability of planetary systems around single stars, but has been modified slightly to include the average semimajor axis between adjacent pairs of planets and the mass interior to the outermost body \citep{Chambers1996}.  Using this formalism, we measure the dynamical fullness of the system.  The mathematical definitions of the mutual Hill Radius, $R_{H,m}$, and dynamical spacing, $\beta$, between the $k,k+1$ planets with identical mass, $m$, are:

\begin{align}
R_{H,m} &= \frac{1}{2}(a_k + a_{k+1})\left(\frac{2m}{3M_T}\right)^{1/3}, \;{\rm and} \\
\beta &= \frac{a_{k+1} - a_k}{R_{H,m}}, \label{eq:beta}
\end{align}
where $M_T = M_A + M_B + m$ and represents the total mass interior to the outermost body.  Our analysis uses this representation to evaluate whether a planet of identical mass could be placed at $a_c$ in addition to the observed CBP.  Others have investigated planet packing for CBPs and determined that values of $\beta = 5 - 7$ would represent the minimum dynamical spacing necessary near the stability limit \citep{Kratter2014,Andrade2017}.  {We use this formalism to measure the dynamical fullness of each system, where dynamically full systems relative to the stability limit are potential evidence for a pile-up in the CBPs.}

\subsection{Effects of the Binary Orbit on the Stability Limit} \label{sec:binary}
We use numerical simulations to identify the stability limit for a Jupiter-mass planet in an initially circular, coplanar orbit around a range of binary parameters.  The results of such simulations can vary with the assumed binary orbit.  Therefore, we first identify the range of variation we can expect when the binary begins at either periastron ($\lambda_{bin}=0^\circ$) or apastron ($\lambda_{bin}=180^\circ$).

\begin{figure}
\centering
\epsscale{1.0}
\includegraphics[width=\linewidth]{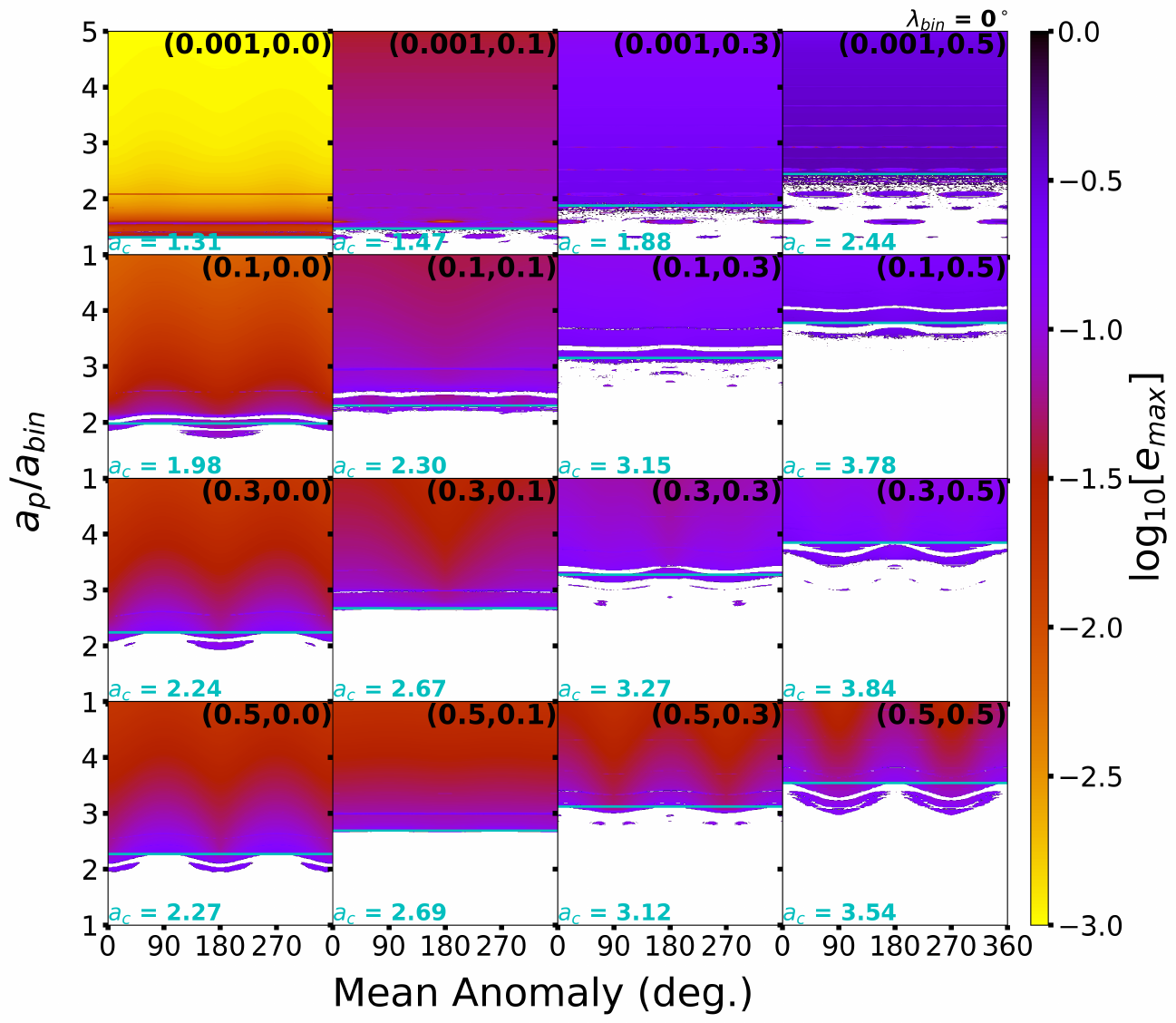}
\caption{Simulations broadly varying the binary parameters ($\mu,e_{bin}$) starting the binary stars at periastron ($\lambda_{bin} = 0^\circ$) and considering the full range of planetary initial Mean Anomalies ($0^\circ - 360^\circ$) in 2$^\circ$ increments.  The color scale represents the maximum eccentricity obtained by a planet over the simulation time ($10^5$ binary orbits), where only stable initial conditions are plotted.  The horizontal cyan lines identify the location of the critical semimajor axis $a_c$ in units of $a_{bin}$, where the respective values are also given.  We note that most structures are symmetric about $180^\circ$ with respect to the planetary Mean Anomaly.} 
\label{fig:bin0}
\end{figure}

Figure \ref{fig:bin0} illustrates how the maximum eccentricity (color-coded) of the planet varies with respect to the initial semimajor axis ratio ($a_p/a_{bin}$) and planetary Mean Anomaly when the binary begins at periastron.  Each subplot varies the binary stellar parameters ($\mu,e_{bin}$), where the respective values are given in the upper right corner.  Additionally the stability limit $a_c$ is identified by a horizontal cyan line and the value is given in the lower left corner.  The white space denotes regions of parameter space that are unstable on the timescale of $10^5$ binary orbits.

The stability limit in the top row of Fig. \ref{fig:bin0} increases as the binary eccentricity, $e_{bin}$, increases.  There are also increases in $a_c$ as $\mu$ increases (i.e., starting from the top row and going down a given column).  However the largest value in $a_c$ does not occur at both the largest $\mu$ and $e_{bin}$ combination, rather when $e_{bin}$ is large (0.5) and $\mu$ is modest (0.1 -- 0.3).  Also, the stability islands--that depend on mean motion resonances--are symmetric about 180$^\circ$ in the planetary Mean Anomaly.  \cite{Deck2013} found similar results when investigating first-order resonance overlap in close two planet systems around a single star, where a similar dynamical environment exists.  The symmetry justifies our choice to investigate only from $0^\circ - 180^\circ$ in the initial planetary Mean Anomaly in the more computationally expensive portion of our study (See Section \ref{results}).

\begin{figure}
\centering
\epsscale{1.0}
\includegraphics[width=\linewidth]{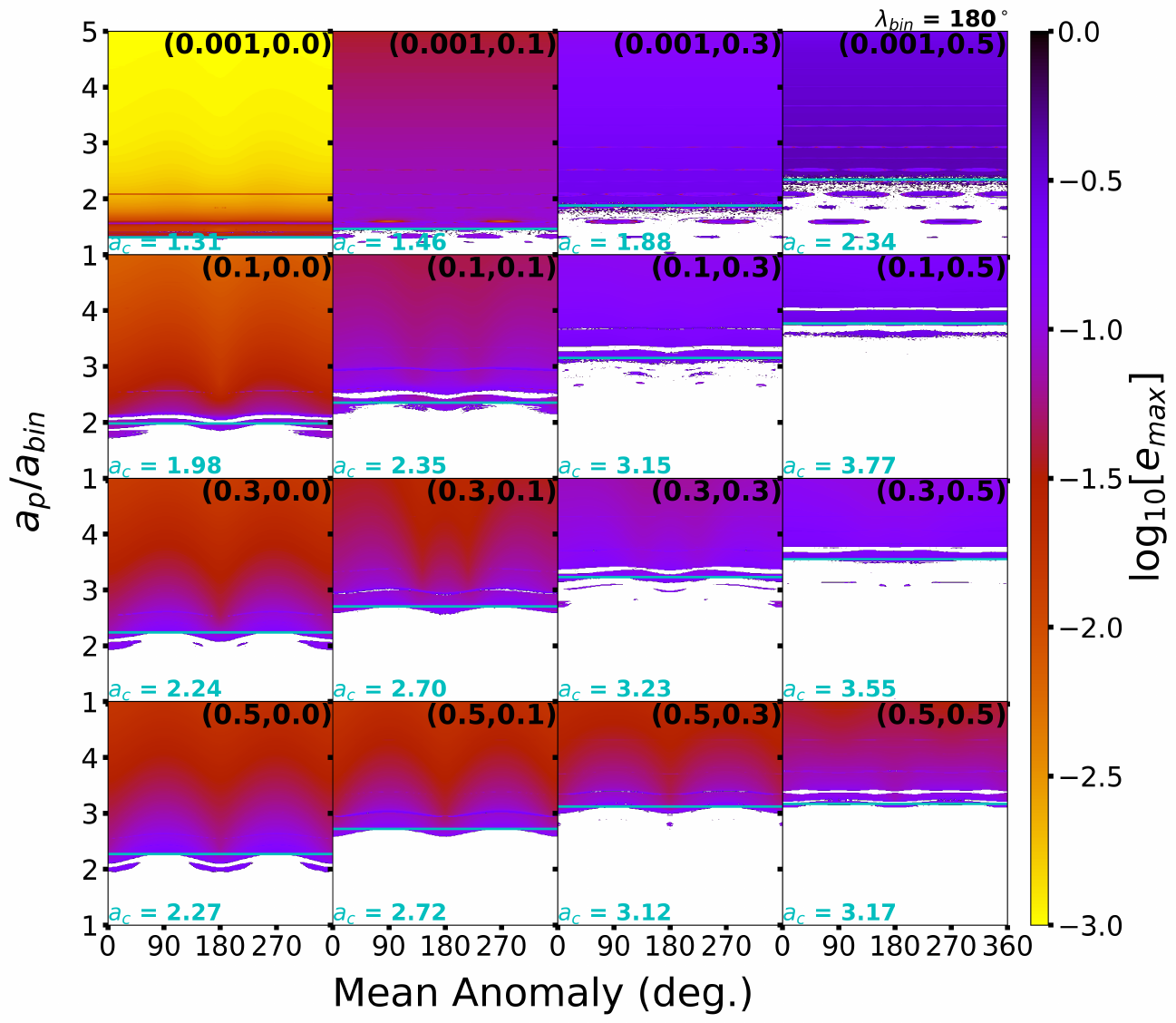}
\caption{Similar to Figure \ref{fig:bin0}, where the stellar components begin at apastron ($\lambda_{bin} = 180^\circ$) instead.  Symmetries about $180^\circ$ with respect to the planetary Mean Anomaly also persist.} 
\label{fig:bin180}
\end{figure}

In contrast, Figure \ref{fig:bin180} demonstrates how the stability limit $a_c$ changes when the binary begins at apastron ($\lambda_{bin}=180^\circ$).  Similar trends are present and most changes in $a_c$ between respective subplots are small ($<$ 0.1).  The most drastic change occurs when $\mu = 0.5$ and $e_{bin} = 0.5$, where the difference in $a_c$ is 0.37. In order to produce conservative (and possibly more reliable) results, we begin the binary at periastron for simulations used in the rest of our analysis.

\section{Results and Discussion} \label{results}
\subsection{Stability Limits Revisited} \label{sec:limits}
We perform a multitude of simulations\footnote{The results of our simulations are publicly available on zenodo.org as a compressed tar archive.  See Section \ref{sec:tools} for details.} ($\sim$150 million) to improve the accuracy of the stability limit, $a_c$, for CBPs.  We determine the $a_c$ for a given combination of binary parameters, $\mu$ and $e_{bin}$, using a grid of simulations (e.g., Figs. \ref{fig:bin0} \& \ref{fig:bin180}), where we limit  initial planetary Mean Anomaly to 180$^\circ$ (see Section \ref{sec:binary}).  

\begin{sidewaysfigure}
\centering
\epsscale{1.0}
\includegraphics[width=\linewidth]{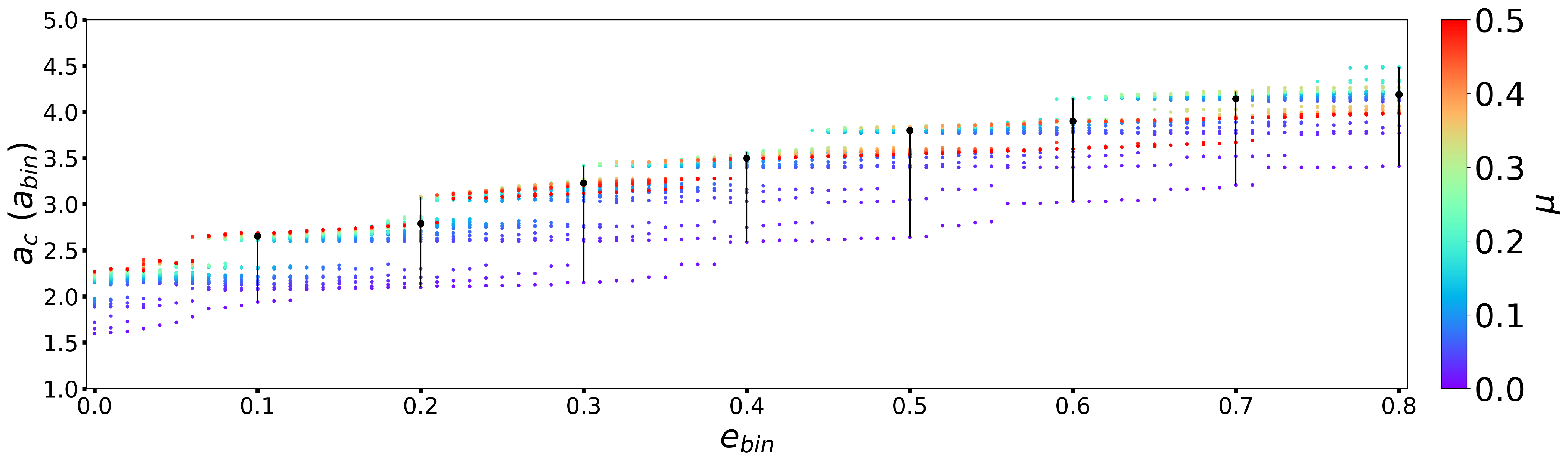}
\caption{Critical semimajor axis $a_c$ values as a function of the binary eccentricity $e_{bin}$, where $\mu \geq 0.01$.  The black points illustrate the median value of $a_c$ in units of $a_{bin}$ over the distribution of $\mu$ considered, where the upper and lower limits show the extreme values attained relative to the median.  There is a large variance in values for $\mu<0.1$. } 
\label{fig:ac_fits}
\end{sidewaysfigure}

From these results, we analyze how $a_c$ varies at a given binary eccentricity, $e_{bin}$, as a function of the binary mass ratio.  Figure \ref{fig:ac_fits} shows these results where the color-code represents the binary mass ratio, $\mu$, and the smallest value $\mu = 0.001$ is excluded because it is significantly flatter relative to the rest of the points.  When including this broad range of mass ratio, the value of $a_c$, can typically vary by $\sim$0.5 -- 1.0 $a_{bin}$.  This variation changes with the binary eccentricity and the median value is not proportional to the mean value.  We have overplotted the median value (black points) with error bars indicating the upper and lower extremes in Fig. \ref{fig:ac_fits} to illustrate how the stability limit is affected by the binary mass ratio, $\mu$.

Most of the variation in the lower bound occurs for $\mu < 0.1$.  If points with $\mu <0.1$ were excluded, then a polynomial function could approximately reflect $a_c$ statistically.  HW99 included results where $\mu = 0.1 - 0.5$ and determined a quadratic polynomial to be appropriate, although their selection on $\mu$ was to exclude a regime that can be modeled using Hill stability.

\begin{figure}
\centering
\epsscale{1.0}
\includegraphics[width=\linewidth]{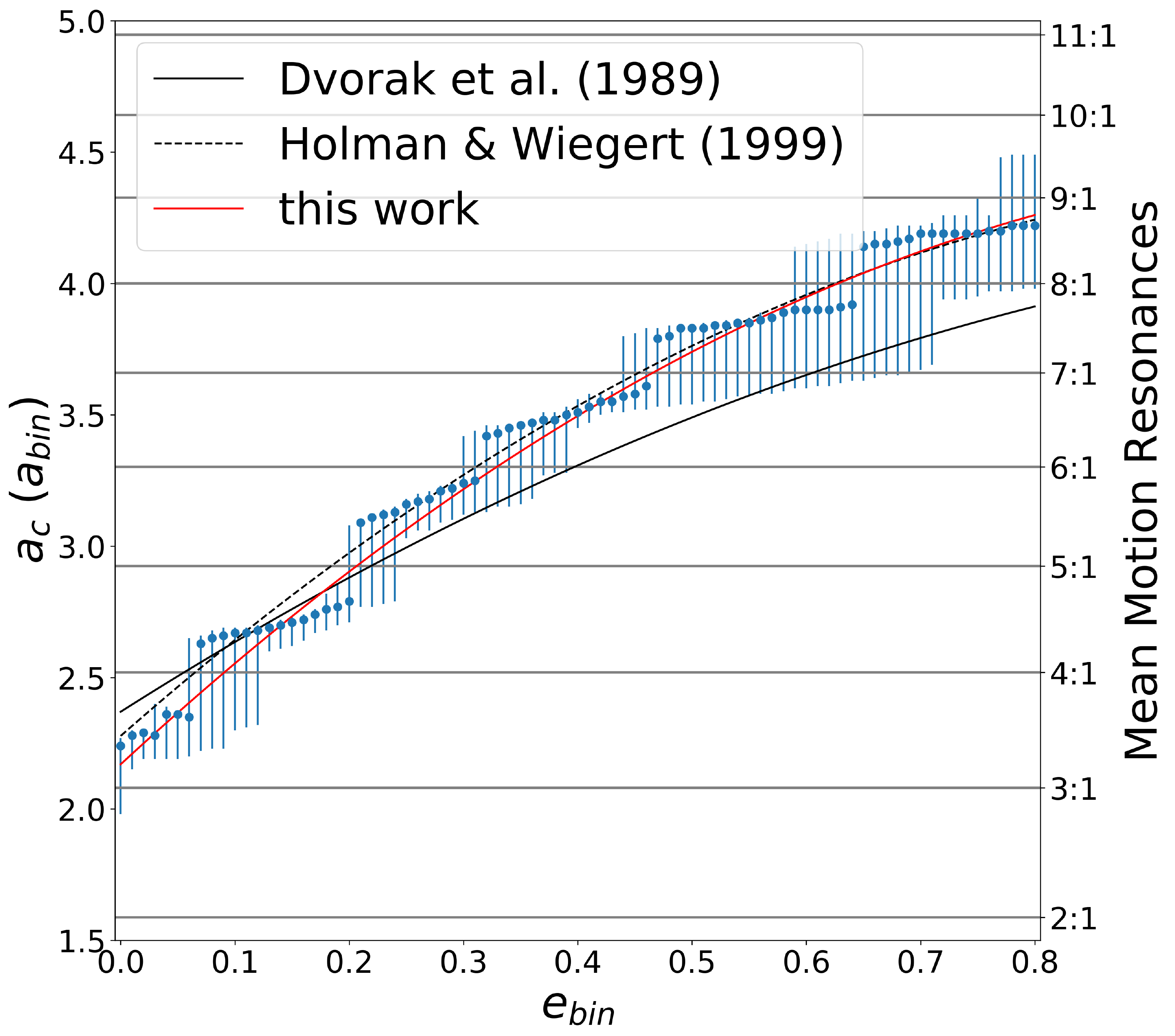}
\caption{Similar to Figure 4 in \cite{Holman1999} considering a quadratic form of $e_{bin}$ after averaging over possible values of $\mu \geq 0.1$.  The blue points indicate the median value, where the upper and lower limits are the extrema values obtain over the distribution of $\mu$ considered.  Three curves are over-plotted showing the relative fit of each work, where the coefficients for each curve are given in Table \ref{tab:Coeff_alt}.  The $N:1$ mean motion resonances are marked along the right y-axis and horizontal (gray) lines are provided to guide the eye.} 
\label{fig:ac_ebin}
\end{figure}

In order to make a fair comparison with HW99, we plot in Figure \ref{fig:ac_ebin} the median values of the stability limit, $a_c$, using error bars to indicate the total range (maximum/minimum values) at a given binary eccentricity $e_{bin}$.  In addition to the data points (blue), we also include the curves using the respective coefficients from \cite{Dvorak1989} (solid black), HW99 (dashed black), and those determined from our simulations (solid red).  The coefficients and uncertainties are provided in Table \ref{tab:Coeff_alt}, where the values for \cite{Dvorak1989} and HW99 are both quoted from HW99 due to the possible errors in labeling noted by HW99.

\begin{deluxetable*}{lcccccccc}
\tablecaption{Coefficients for the Critical Semimajor Axis Using $e_{bin}$ \label{tab:Coeff_alt}}
\tablecolumns{4}
\tablehead{\colhead{} & \colhead{$C_1$} & \colhead{$C_2$} & \colhead{$C_3$} } 
\startdata
\cite{Dvorak1989} & 2.37 & 2.76 & -1.04 \\
HW99 & 2.278$^{+0.008}_{-0.008}$ & 3.824$^{+0.33}_{-0.33}$ & -1.71$^{+0.10}_{-0.10}$ \\
this work & 2.170$^{+0.017}_{-0.017}$ & 4.017$^{+0.10}_{-0.10}$ & -1.75$^{+0.14}_{-0.14}$ \\
\enddata
\tablecomments{The coefficients (and uncertainties) for $C_1$, $C_2$, and $C_3$ from previous studies are listed that use a quadratic fitting function ignoring $\mu$, $a_c/a_{bin} = C_1 + C_2e_{bin} +  C_3e_{bin}^2$.  We use the same function in this work but use the maximum, median, and minimum values of $a_c$ (e.g., Fig. \ref{fig:ac_ebin}).}
\end{deluxetable*}

Upon inspection of Fig. \ref{fig:ac_ebin} and Table \ref{tab:Coeff_alt}, we reaffirm the previous results, where most of our coefficients overlap (within errors) with those of HW99.  However, both fits are applicable at a statistical distribution level and not very accurate individually due to the effective marginalization over the binary mass ratio, $\mu$.  In Fig. \ref{fig:ac_ebin}, we also mark the expected locations of the mean motion resonances the planet would encounter with the binary orbit{, which act to destabilize CBPs \citep{Mudryk2006}}.  \cite{Doolin2011}, \cite{Quarles2016}, and others have shown through large parameter space studies that these resonances produce unstable gaps and stability islands can exist at locations approximately half-way between the resonances.  

Another method utilized by HW99 is to allow both the binary semimajor axis, $\mu$, and the binary eccentricity, $e_{bin}$, to vary as quadratic functions.  We perform a similar approach (using all our data) and provide our results alongside those determined by HW99 in Table \ref{tab:Coeff}.  The reduced chi-square statistic is provided using the HW99 coefficients as well as our own.  We provide an additional fitting where we make the replacement of $\mu \rightarrow \mu^{1/3}$ as motivated by stability studies using the Hill radius (i.e., planet packing around a single star).  Both of our fittings produce a lower chi-square statistic than HW99, although our Fit 2 is likely to be biased in that a large portion of our simulations ($\sim$40\%) have a binary mass ratio within the Hill regime.

\begin{deluxetable*}{lcccccccc}
\tablecaption{Coefficients for the Critical Semimajor Axis \label{tab:Coeff}}
\tablecolumns{9}
\tablehead{\colhead{} & \colhead{$C_1$} & \colhead{$C_2$} & \colhead{$C_3$} & \colhead{$C_4$} & \colhead{$C_5$} & \colhead{$C_6$} & \colhead{$C_7$} & \colhead{$\chi^2_\nu$}} 
\startdata
HW99 & 1.60$^{+0.04}_{-0.04}$ & 5.10$^{+0.05}_{-0.05}$ & -2.22$^{+0.11}_{-0.11}$ & 4.12$^{+0.09}_{-0.09}$ & -4.27$^{+0.17}_{-0.17}$ & -5.09$^{+0.11}_{-0.11}$ & 4.61$^{+0.36}_{-0.36}$ & 2015.97\tablenotemark{a}\\
Fit 1 & 1.48$^{+0.01}_{-0.01}$ & 3.92$^{+0.06}_{-0.06}$ & -1.41$^{+0.06}_{-0.06}$ & 5.14$^{+0.10}_{-0.10}$ & 0.33$^{+0.19}_{-0.19}$ & -7.95$^{+0.15}_{-0.15}$ & -4.89$^{+0.44}_{-0.44}$ & 876.25\\
Fit 2 & 0.93$^{+0.02}_{-0.02}$ & 2.67$^{+0.08}_{-0.08}$ & -0.25$^{+0.06}_{-0.06}$ & 3.72$^{+0.06}_{-0.06}$ & 2.25$^{+0.12}_{-0.12}$ & -2.72$^{+0.05}_{-0.05}$ & -4.17$^{+0.15}_{-0.15}$ & 450.55\tablenotemark{b}\\
\enddata
\tablecomments{The coefficients and uncertainties for $C_1 - C_7$ from HW99 are listed using the fitting formula,  $a_c/a_{bin} = C_1 + C_2e_{bin} +  C_3e_{bin}^2 + C_4\mu + C_5e_{bin}\mu + C_6\mu^2 + C_7e_{bin}^2\mu^2$.  We perform two separate fits (Fit 1 and Fit 2) using all our data and list the resulting reduced chi-square value, $\chi^2_\nu$.}
\tablenotetext{a}{This value was calculated using the coefficients listed in HW99 and our larger dataset.}
\tablenotetext{b}{This fit modifies the equation where $\mu \rightarrow \mu^{1/3}$ in order to better match the form with the Hill radius when $e_{bin}$ and $\mu$ are small.}
\end{deluxetable*}

Although we find good agreement statistically with HW99, the final result will represent CBPs at the population level and there are not enough detections made thus far to justify a completely statistical treatment.  Therefore, we suggest a different approach, which is to think of a stability surface (i.e., two-dimensional) rather than a stability limit.  In this interpretation, we can obtain much higher accuracy at a individual system level through grid interpolation of our results.  

Figure \ref{fig:full_space} illustrates how our dataset can be used to make such a map\footnote{We provide python tools on GitHub to query our dataset and reproduce all of our figures.  Specifically, there is a routine that returns $a_c$ through grid interpolation for a given combination of $\mu$ and $e_{bin}$.  See Section \ref{sec:tools} for details}.  Each point is color-coded to the stability limit, $a_c$, determined through a smaller grid of simulations (e.g., Fig. \ref{fig:Kepler_ac}).  Additionally, in Fig. \ref{fig:full_space}, we have over-plotted (white dots) the locations corresponding to the stellar parameters of the Kepler CBPs.  The smallest value of $a_c$ is 1.31 and is located where one would expect.  Interestingly, the largest value of $a_c$ is 4.49 and is not produced considering the largest value of $\mu$ that we consider.  HW99 also observed a similar feature ($a_c = 4.2 - 4.3$), but their range in $e_{bin}$ and resolution did not allow them to identify this location accurately.

\subsection{Comparison to the Kepler CBPs -- Populations} \label{sec:pop}
 The Kepler mission has uncovered 9 CBP systems, whose stellar and planetary properties vary widely and comparing them statistically in terms of a stability limit may not be reliable.  \cite{Li2016} examined how the mutual inclination of CBPs relative to the binary orbital plane would affect the probability of observing a pile-up of the Kepler CBPs at the stability limit.  Their study demonstrated that different conclusions can be drawn when Kepler-1647 is and is not included in the sample of CBPs and more systems need to be observed in order to distinguish between their two scenarios.
 
As a result, we compare each of the Kepler CBPs at a system-by-system level using the values of the critical semimajor axis, $a_c$ determined in Section \ref{sec:limits}.  We also note that our analysis represents a conservative estimate of the stability limit as our determined limits for $a_c$ could decrease with an increased mutual inclination of the CBP \citep{Doolin2011,Li2016} or the stellar binary begins closer to apastron rather than periastron (See Section \ref{sec:binary}). Table \ref{tab:Star_param} summarizes the observationally determined stellar masses and orbital parameters of each of the known Kepler CBPs.

\begin{deluxetable*}{lcccccccc}
\tablecaption{Stellar Parameters for the Kepler CBPs \label{tab:Star_param}}
\tablecolumns{9}
\tablehead{\colhead{} & \colhead{$M_A$ ($M_\odot$)} & \colhead{$M_B$ ($M_\odot$)} & \colhead{$\mu$} & \colhead{$a_{bin} (AU)$} & \colhead{$e_{bin}$} & \colhead{$\omega$ (deg.)} & \colhead{$MA$ (deg.)} & Ref.} 
\startdata
Kepler-16 & 0.6897 & 0.20255 & 0.2270 & 0.22431 & 0.15944 & 263.464 & 188.888 & \cite{Doyle2011} \\
Kepler-34 & 1.0479 & 1.0208 & 0.4934 & 0.22882 & 0.52087 & 71.437 & 228.760  & \cite{Welsh2012} \\
Kepler-35 & 0.8877 & 0.8094 & 0.4769 & 0.17617 & 0.1421 & 89.1784 & 2.9021 & \cite{Welsh2012} \\
Kepler-38 & 0.949 & 0.249 & 0.208 & 0.1469 & 0.1032 & 268.68 & 181.32  & \cite{Orosz2012a} \\
Kepler-47 & 0.957 & 0.342 & 0.263 & 0.08145 & 0.0288 & 226.253 & 310.818  & \cite{Orosz2012b} \\
Kepler-64 & 1.528 & 0.408 & 0.211 & 0.1744 & 0.2117 & 219.7504 & 251.558  & \cite{Schwamb2013} \\
Kepler-413 & 0.820 & 0.5423 & 0.398 &0.10148 & 0.0365 & 279.54 & 169.5328  & \cite{Kostov2014} \\
Kepler-453 & 0.944 & 0.1951 & 0.171 &0.185319 & 0.0524 & 263.05 & 187.7059 & \cite{Welsh2015} \\
Kepler-1647 & 1.2207 & 0.9678 & 0.4422 &0.1276 & 0.1602 & 300.5442 & 139.0749  & \cite{Kostov2016} \\
\enddata
\tablecomments{The stellar parameters ($M_A$, $M_B$, $\mu$, $a_{bin}$, $e_{bin}$, $\omega$, and $MA$) of the Kepler CBPs are listed.  The definitions of these orbital parameters carry their usual meaning {from the exoplanet literature}.}
\end{deluxetable*}

To measure the proximity of the Kepler CBPs to our determined stability limit, we first determine $a_c$ through a grid interpolation of Fig. \ref{fig:full_space} using the $\mu$ and $e_{bin}$ values given in Table \ref{tab:Star_param}.  The result of this interpolation for each CBP is given in Table \ref{tab:pl_param}.  The observed planetary semimajor axis $a_p$ is also provided along with a measure of the percentage difference between $a_p$ and $a_c$.  The comparison using percent difference shows that some CBPs are much closer to $a_c$ than others, where the average difference between $a_p$ and $a_c$ is $\sim$42\%.  For systems that are much lower than 42\%, we initially classify to be at the stability limit and those that are much higher are not at the stability limit.

Another method for determining the proximity to the stability limit uses formalisms from planet packing studies \citep[e.g.,][]{Kratter2014} and require the calculation of the mutual Hill radius, R$_{H,m}$ (See Section \ref{sec:Hill}).  For this calculation, we propose that planets classified to reside at the stability limit should not allow for an additional equal-mass planet to exist on an interior orbit at our determined $a_c$.  Along with R$_{H,m}$, we also determine the dynamical spacing, $\beta_c$, between an equal-mass planet at $a_c$ relative to the observed planet at $a_p$ in Table \ref{tab:pl_param}.  

\cite{Kratter2014} determined that stability is possible with $\beta_c = 5 - 7$, where we define in this analysis that $\beta_c \leq 7$ does not allow for an interior equal-mass planet to exist at $a_c$.  Using this criterion, we find that 5 out of 9 CBP systems (55\%) do allow for an interior equal-mass planet.  {However, the previous study \citep{Kratter2014} did not take the binary eccentricity into account and we perform a limited suite of N-body simulations to confirm the above estimate for the Kepler CBPs.}

{In these simulations, we introduce an equal-mass planet with a semimajor axis between $a_{bin}$ and $q_p (= a_p(1 - e_p))$ with steps of 0.001 AU.  The binary can induce a forced eccentricity on the inner planet \citep[e.g.,][]{Mudryk2006}.  As a result, we choose the initial eccentricity vectors of both planets to be aligned ($\omega = 0^\circ$) with the binary orbit and vary the magnitude of the eccentricity vector from 0.0 - 0.50 in steps of 0.01.  We follow a similar relative phase setup from \cite{Gladman1993}, where each planet pair starts 180$^\circ$ out of phase from one another and the inner planet begins at periastron ($MA = 0^\circ$).  In order to identify robust regions of stability, these simulations are integrated up to 500 million orbits for a planet at $a_c$ (see Table \ref{tab:pl_param} for values of $T_c$).  The integration step is adjusted for each simulation at 2.5\% of the initial Keplerian period for the inner planet.}

\begin{figure}
\centering
\includegraphics[width=\linewidth]{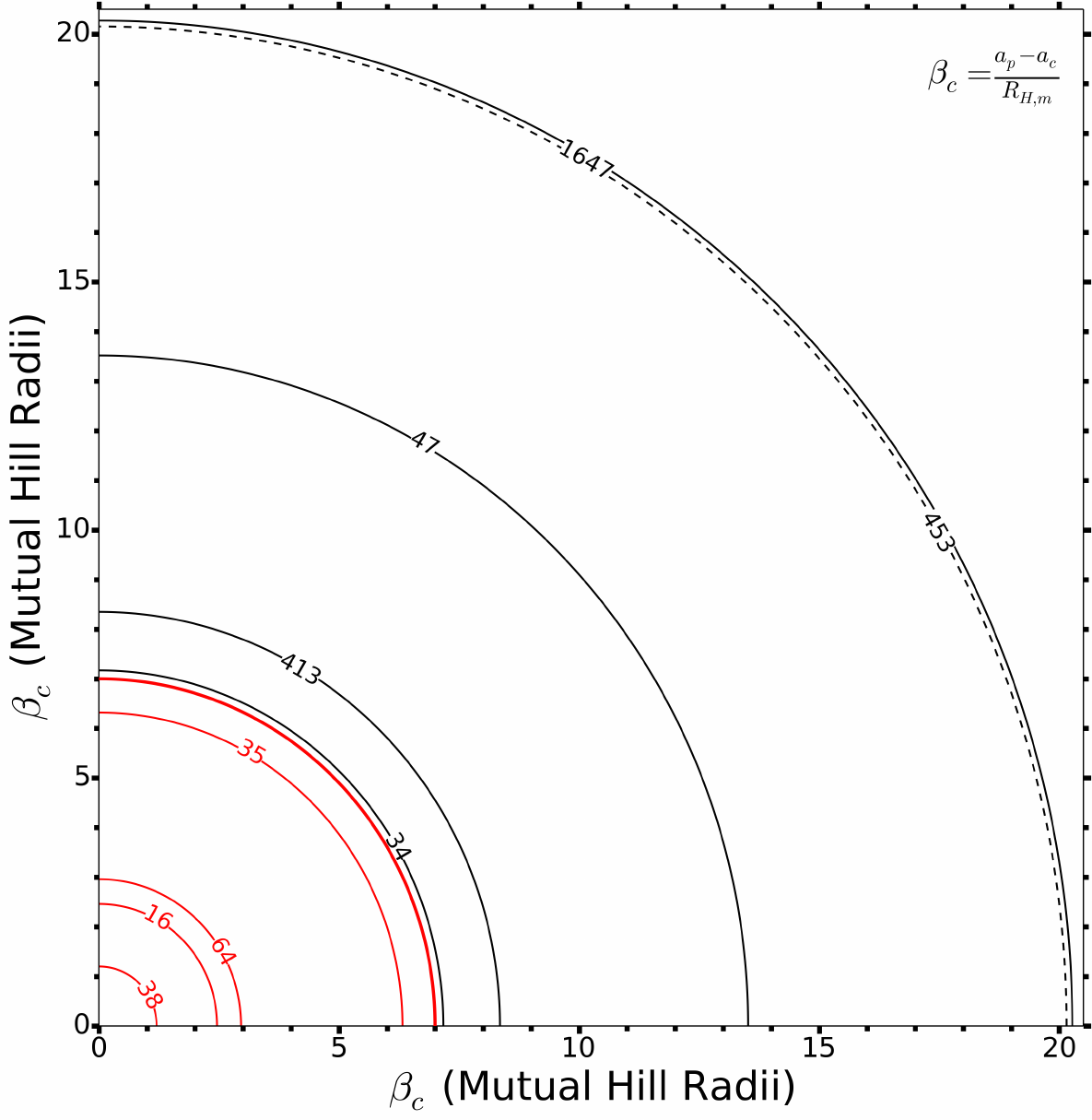}
\caption{{Schematic illustrating the relative difference between the Kepler CBPs using the dynamical spacing, $\beta_c$, between the observed planet at $a_p$ and an equal-mass planet placed at the stability limit, $a_c$. {The red curves designate dynamical spacing values that do \textit{not} allow for stable planets with $a<a_p$ from our numerical simulations and the value ($\beta_c = 7$) from previous works.}  The dynamical spacings for Kepler-453b and Kepler-1647b are delineated by dashed and solid curves, respectively.  See Table \ref{tab:pl_param} for more precise values.}} 
 \label{fig:beta_space}
 \end{figure}

{Our full N-body simulations justify our criterion, $\beta_c > 7$, to allow for additional equal-mass planets to stably orbit within 5 of the Kepler CBP systems.  Figure \ref{fig:beta_space} illustrates the relative distribution of the Kepler CBPs through their respective values of $\beta_c$ as concentric circles, where the origin denotes the location that is exactly at the stability limit.  In this schematic, the dynamical separation, $\beta_c$, from the stability limit does \emph{not} appear to cluster at any particular value.  If we choose the inner planet mass to be Earthlike, then our values of $\beta_c$ would increase by $\sim$2$^{1/3}$ and potentially allow for an additional planet in Kepler-35.  Note: We emphasize that we are \emph{not} confirming the existence of any planets interior to the observed Kepler CBPs.  If such planets do exist, then they must be on sufficiently inclined orbits at the present epoch to have avoided detection.}

\begin{deluxetable*}{lcccccc}
\tablecaption{Stability Limits for the Kepler CBPs \label{tab:pl_param}}
\tablecolumns{7}
\tablehead{\colhead{} & \colhead{$a_{c}$ (AU)} & \colhead{$T_c$ (days)} & \colhead{$a_p$ (AU)} & \colhead{\% diff} & \colhead{R$_{H,m}$ (AU)} & \colhead{$\beta_{c}$} }
\startdata
Kepler-16b & 0.6050 & 182.0 & 0.7048 & 15.24 & 0.0405 & 2.4610 \\
Kepler-34b & 0.8118 & 185.7 & 1.0896 & 29.220 & 0.0387 & 7.1703 \\
Kepler-35b & 0.4795 & 93.09 & 0.6035 & 22.89 & 0.0196 & 6.3175 \\
Kepler-38b & 0.4328 & 95.02 & 0.4644 & 7.047 & 0.0264 & 1.1968 \\
Kepler-47b & 0.1848 & 25.46 & 0.2956 & 46.13 & 0.00820 & 13.519 \\
Kepler-64b & 0.5368 & 103.2 & 0.634 & 16.6  & 0.0327 & 2.9697\\
Kepler-413b & 0.2389 & 36.54 & 0.353 & 38.6 & 0.0136 & 8.3487 \\
Kepler-453b & 0.4184 & 92.62 & 0.7903 & 61.53  & 0.0184 & 20.152 \\
Kepler-1647b & 0.3497 & 51.06 & 2.72 & 154  & 0.117 & 20.275 \\
\enddata
\tablecomments{Calculated values of $a_c$, $T_c$, $\% \;{\rm diff}$, R$_{H,m}$ and $\beta_c$ are listed for each of the Kepler CBPs, where the $a_p$ values are drawn from the discovery papers (see Table \ref{tab:Star_param}).  We use the definition of percent difference as $\% \;{\rm diff} = 2|a_p - a_c|/(a_p + a_c)$.}

\end{deluxetable*}

\subsection{Comparison to the Kepler CBPs -- Individual Systems}
The Kepler CBPs are a snapshot of the larger population of CBPs, where we want to investigate them at a system-by-system level.  We identify the ways that the initial conditions could alter our determination of the critical semimajor axis ratio, $a_c$, by choices in the: initial phase of the binary orbit, initial phase of the planetary orbit, or the initial eccentricity of the planetary orbit.

We examine the stability of systems \textit{similar} to the Kepler CBPs in the binary parameters ($\mu,e_{bin}$) using the results from our simulations in Fig. \ref{fig:full_space}.  Figure \ref{fig:Kepler_ac} illustrates the variation of stability over $10^5$ binary orbits with respect to variations in the initial semimajor axis ratio and Mean Anomaly of the planetary orbit, when the binary begins at periastron ($\lambda_{bin}=0^\circ$).  We emphasize these assumptions because the actual Kepler CBPs will likely not adhere to them and shifts in the initial phase may be necessary for 1:1 comparisons. 

In Fig. \ref{fig:Kepler_ac}, most systems are not strongly dependent on the choice of the planetary Mean Anomaly (except Kepler-34, Kepler-38, and Kepler-64) and our definition of stability (see Section \ref{sec:stab}) appears to be robust when stability islands exist at specific ranges in the Mean Anomaly of the planetary orbit.  Comparing the values of $a_c$ at these points are consistent (within $\sim$1\%) to those given in Table \ref{tab:pl_param}, after multiplying by $a_{bin}$, that were determined through a grid interpolation.  

\begin{sidewaysfigure}
\centering
\includegraphics[width=\linewidth]{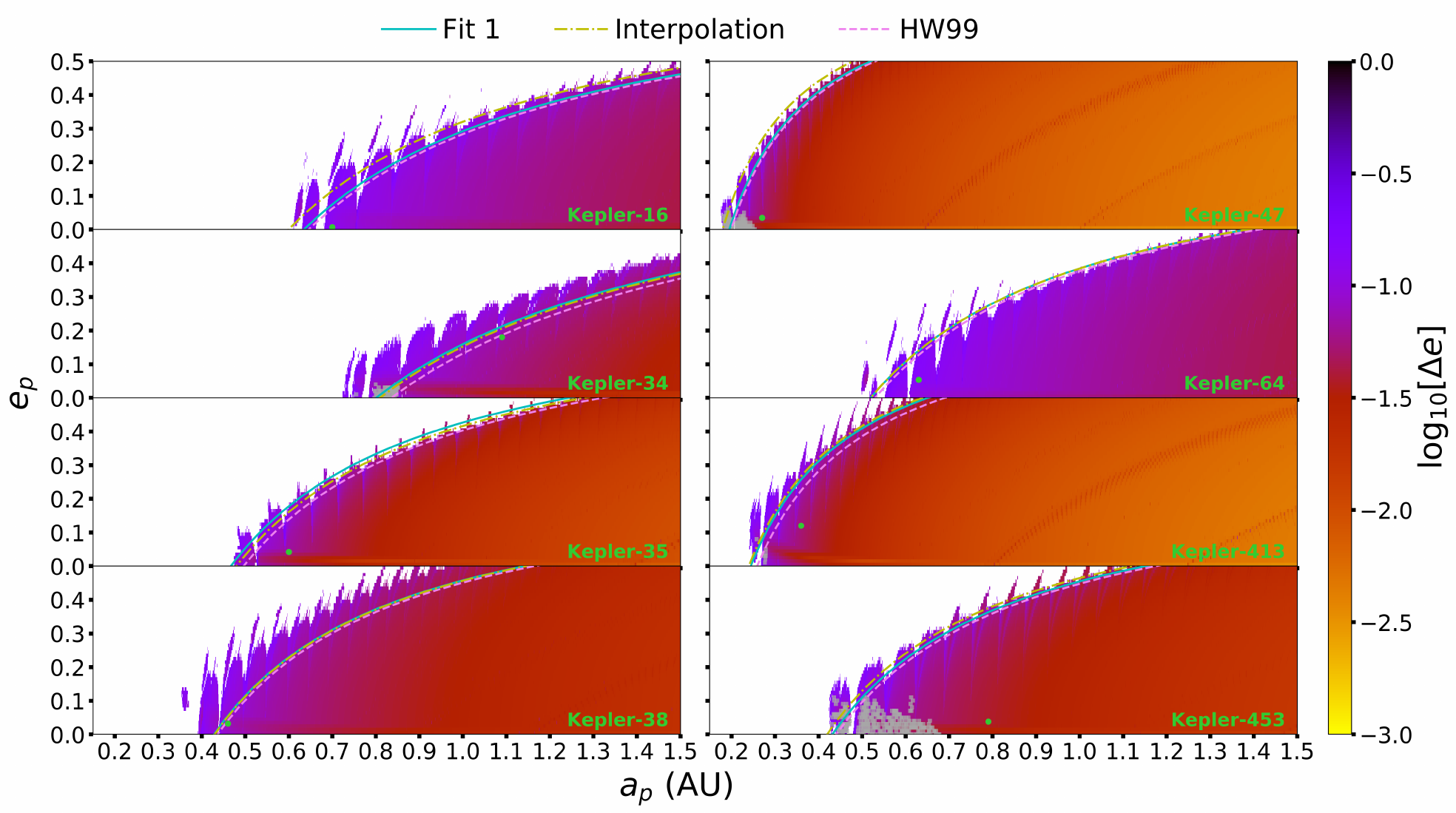}
\caption{Stability of a Jupiter-mass planet orbiting the Kepler CBP host stars considering a wide range of initial conditions with respect to the planetary semimajor axis, $a_p$, and planetary eccentricity, $e_p$.  The planet begins oriented along the reference node ($\omega = \Omega = MA = 0^\circ$), while the host binary begins with the parameters listed in Table \ref{tab:Star_param}.  The color-code represents the range of eccentricity ($\Delta e = e_{max} - e_{min}$) attained over the simulation time for stable initial conditions on a base-10 logarithmic scale, where only stable initial conditions are plotted.  The curves denote the approximate stability limit assuming that $a_c \propto q_c$ using our coefficients in Fit 1 (cyan, see Table \ref{tab:Coeff})), our interpolated values (yellow, see Table \ref{tab:pl_param}), and the coefficients from HW99 (violet, see Table \ref{tab:Coeff}).  The green dots illustrate the location of the observed CBPs within this parameter space.  {The gray squares demonstrate regions of the parameter space where an equal-mass planet (in addition to the Kepler CBP) can remain stable up to 500 million orbits.}} 
 \label{fig:Kepler_stab}
 \end{sidewaysfigure}

We go beyond our ideal setup (excluding Kepler-1647, see \cite{Kostov2016} for a stability map) that makes assumptions on the binary and planetary orbit.  For this, we evaluate the variation of stability considering the actual host binary orbit (see Table \ref{tab:Star_param}) with a range of initial eccentricity (0 -- 0.5 in steps of 0.01) and semimajor axis ($a_{bin}$ -- 1.5 AU in steps of 0.001 AU) for a Jupiter-mass planet.  The planet begins along the reference node so that $\omega = \Omega = MA = 0^\circ$.  We plot the initial conditions that are stable for at least 100,000 years in Figure \ref{fig:Kepler_stab} using a color-code, the location (green dot) of the observed Kepler CBP parameters, and over-plot the approximate stability boundary for 3 methods: our Fit 1 (cyan, see Table \ref{tab:Coeff}), our interpolation (yellow, see Table \ref{tab:pl_param}), and HW99 (violet, see Table \ref{tab:Coeff}) using mean values where applicable.  The stable initial conditions are color-coded based upon the range of eccentricity ($\Delta e = e_{max}-e_{min}$) a planet attains over the simulation time on a base-10 logarithmic scale.  \cite{Ramos2015} and \cite{Giuppone2017} have used a similar metric because it highlights dynamical regions affected by resonant interactions, where our definition differs from theirs by a factor of 2 for clarity.  Our results from Section \ref{sec:pop} investigating whether interior planets could be possible are also shown as gray squares, where those simulations used planet pairs more similar to the actual mass of the Kepler CBPs.

For the stability boundary, we assume that a critical pericenter distance, $q_c$, exists for an eccentric orbit that corresponds approximately to the critical semimajor axis for a circular orbit \citep[i.e.,][]{Popova2016}.  The mathematical expression that we use to approximate the critical eccentricity, $e_c$, is:

\begin{align}
e_c &= 0.8\left(1 - \frac{a_c}{a_p}\right),
\end{align}

where $a_c$ is the critical semimajor axis (in AU) derived via each method, and $a_p$ is the planetary semimajor axis in AU.  The 0.8 factor in our equation is arbitrary, but we found that using this value consistently improves the fit of the upper boundary (high values of $e_p$) of stability for most cases.  Our interpolation method (yellow curve) in Fig. \ref{fig:Kepler_stab} typically agrees well with the innermost stable circular orbit (within $\sim$1\%).  The other 2 methods (Fit 1 \& HW99) also agree within their error limits, although a substantial fraction within the error range exists in a region of unstable parameter space.

\subsection{CBPs in Context: Observations with TESS}

With a sample of only 10 planets, one of which is already an outlier in terms of orbital separation (Kepler-1647b), interpretations of the available data--such as the proposed pile-up of CBPs at the dynamical stability limit--may be affected by observational bias. Increasing the sample by a factor of two would be very useful.  Increasing it by a factor of ten would be fantastic and, with the help of the tools we develop here, would enable comprehensive statistical studies for or against the potential pile-up. 

An order of magnitude increase in the number of known CBPs will be indeed possible with the TESS mission by using a novel method for CBP detection based on the occurrence of multiple transits during the same conjunction. This method has already been demonstrated for the case of two such transits of Kepler-1647b, where \cite{Kostov2016} estimated a planet period within 5\% of the true period by combining radial velocity measurements with transit timing--independently from the full photodynamical solution of the systems. Similar transits can easily occur within the 30-day all-sky observing window of TESS.

TESS will observe the entire sky for at least 30 days, and continuously measure the brightness of $\sim$20 million stars (including $\sim$500,000 eclipsing binaries) brighter than $R\sim15$ with mmag precision \citep{Sullivan2015}.  Based on the CBP results from Kepler, we expect the CBP yield of TESS to be a few hundred planets similar to Kepler's (Kostov et al in prep). The tools and methods we describe here will be directly applicable for both estimating the stability of each new CBP candidate during the initial detection phase, as well as for detailed dynamical investigations of the entire sample after the comprehensive photodynamical characterization of all planets.  For example, our method would allow rapid identification ($<$1 second) of the likelihood that a candidate is a false positive--and thus immediately guide follow-up efforts--based on stability criteria and dynamical packing. In addition, if any TESS CBP system exhibits extra transits, not associated with either the binary or the detected planets, by applying the methodology presented here we will be able to rule out the orbital parameter space available to additional planets in the respective systems.

{\subsection{Numerical Tools for the Community} \label{sec:tools}
We perform a multitude of simulations into order to determine the most general and reliable stability limit given a set of binary parameters ($\mu$, $e_{bin}$).  The results of these simulations are available through \texttt{GitHub}\footnote{\url{https://github.com/saturnaxis/CBP_stability}} and \texttt{Zenodo}\footnote{\url{http://doi.org/10.5281/zenodo.1174228}}.  The \texttt{GitHub} repository contains scripts to identify the stability limit, $a_c$, and reproduce the figures contained in this paper using \texttt{Matplotlib} \citep{Hunter2007,Droettboom2016}.  The determination of $a_c$ is not limited to the binary parameters used to make Fig. \ref{fig:full_space}, but can be interpolated using routines from \texttt{Scipy} \citep{Jones2001} in Python or other programming languages \citep{Press1992}.}

{The full dataset is available as a compressed tar archive on \texttt{Zenodo}.  The archive contains text files that are delineated by the assumed binary parameters ($\mu$, $e_{bin}$) in the filenames.  Python scripts to manipulate the dataset without extracting all the files are available in the \texttt{GitHub} repository.  Each comma delimited file in the archive lists the results of a given simulation, where the columns are the initial semimajor axis ratio, the initial planetary phase in degrees, the maximum planetary eccentricity attained, the minimum eccentricity attained, and the collision/escape time in years.  For initial conditions that survived the full simulation time, a value of $10^5$ yr is reported the final column.}

\section{Conclusions} \label{sec:conc}
The number of known circumbinary planets (CBPs) is currently small ($\sim$10), but the current methods to determine the proximity of these CBPs to the stability limit for their host stars is statistical (HW99).  In this paper, we perform a multitude of numerical simulations ($\sim$150 million) to better understand the stability surface of CBPs as a function of stellar mass ratio, $\mu$, and eccentricity, $e_{bin}$.  We provide open-source python software for the community to access and make use of our simulations, specifically with a python script that can interpolate our results for the \textit{stability surface} for CBP candidates derived from photometric planet surveys.

\begin{figure}
\centering
\epsscale{1.0}
\includegraphics[width=\linewidth]{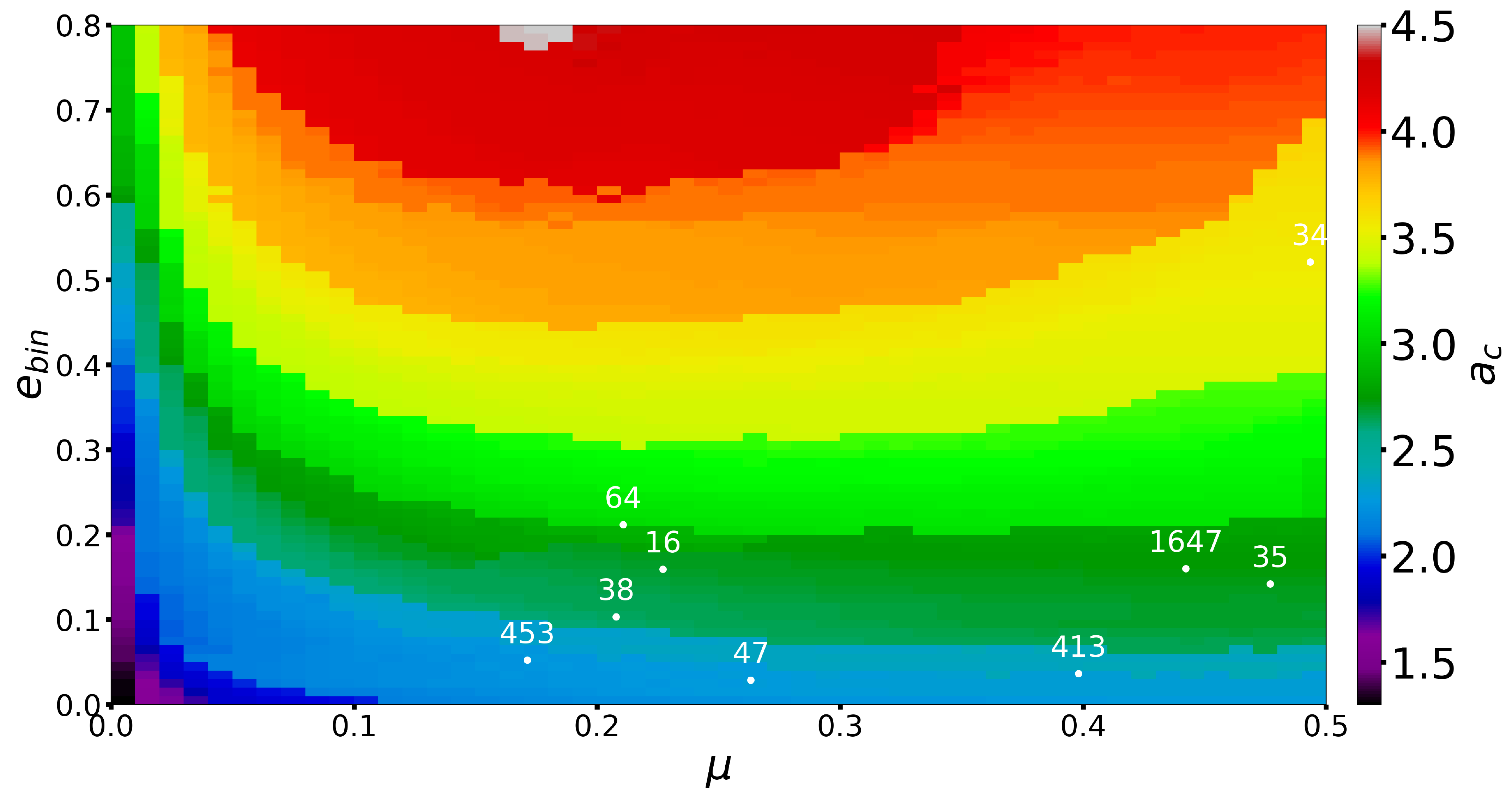}
\caption{Smallest stable planetary semimajor axis $a_c$ in units of $a_{bin}$ over all initial planetary Mean anomalies as a function of the binary mass ratio $\mu$ and eccentricity $e_{bin}$.  The white dots represent the corresponding locations of the host stars for the Kepler CBPs within this parameter space.} 
\label{fig:full_space}
\end{figure}

\begin{figure}
\centering
\includegraphics[width=\linewidth]{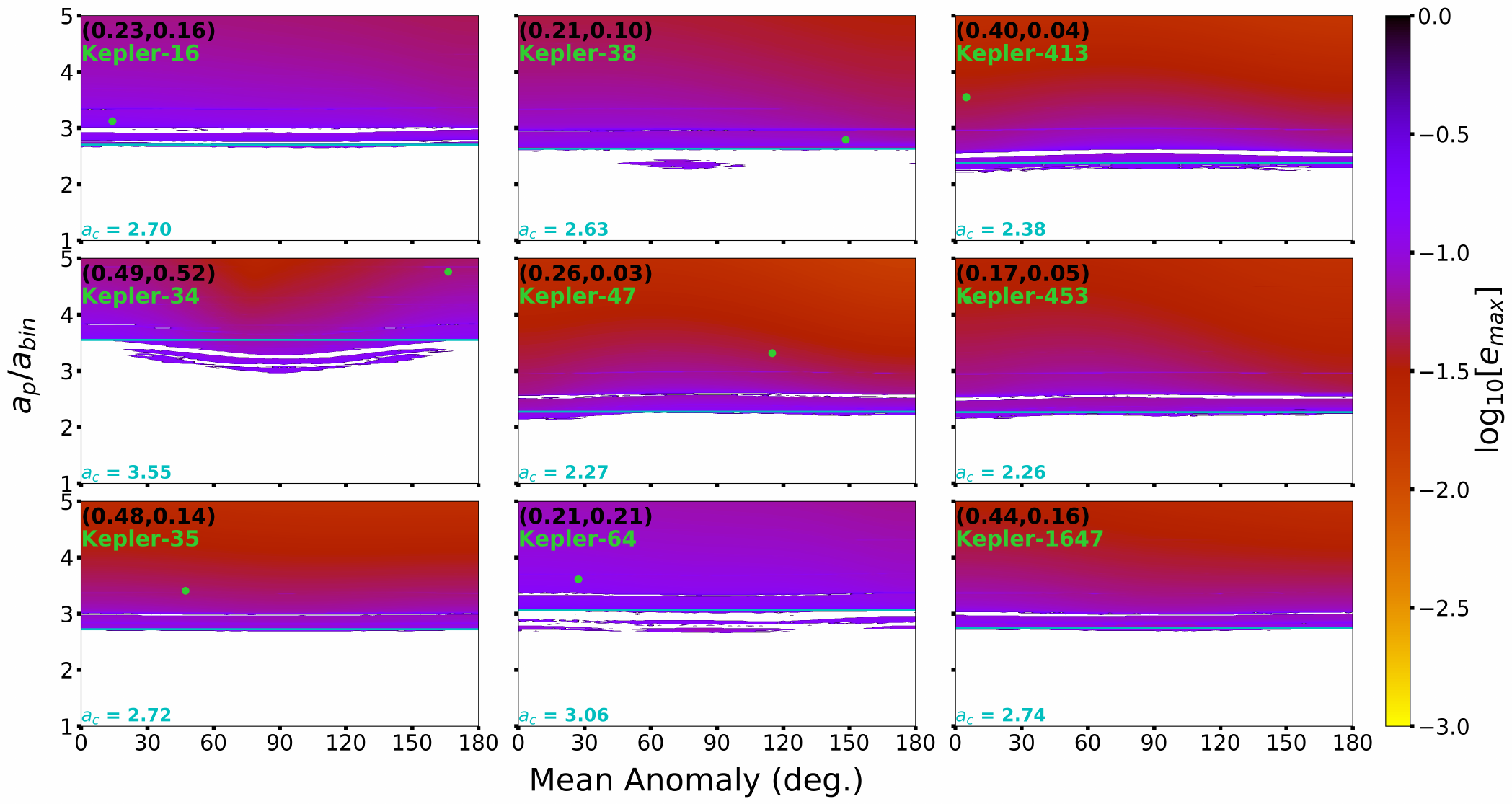}
\caption{Stability maps of an initially coplanar, circular Jupiter-mass planet using our dataset ($\sim$150 million simulations) in Fig. \ref{fig:full_space} at the nearest combination of stellar parameters ($\mu,e_{bin}$).  Similar to Fig. \ref{fig:bin0}, the stellar binary begins at periastron ($\lambda_{bin} = 0^\circ$), while the initial semimajor axis ratio and Mean Anomaly of the planetary orbit are varied. {These simulations cover a wide range of initial conditions that include configurations consistent with the architecture of the Kepler CBPs (green dots).  The semimajor axis of the CBPs in Kepler-453 and Kepler-1647 are more than 5 times $a_{bin}$ and are not plotted.}  The color scale represents the maximum eccentricity of the planet attained over $10^5$ binary orbits on a base-10 logarithmic scale, where only stable initial conditions are plotted.  The horizontal cyan lines identify the location of the critical semimajor axis $a_c$ in units of $a_{bin}$, where the respective values are also given. } 
 \label{fig:Kepler_ac}
 \end{figure}

Using our numerical tools, we devise a grid interpolation method that uses the stability surface to accurately characterize the inner limits of stability with respect to $\mu$ and $e_{bin}$ (see Figure \ref{fig:full_space}).  We compare our derived stability limits to the previous study by \cite{Holman1999} for completeness and find good agreement, within errors.  The reduced chi-square of our fits are smaller than HW99, which is likely a result of the increased resolution.  We find that replacing $\mu \rightarrow \mu^{1/3}$ provides a better fit due to the weak dependence on the stellar mass ratio \citep{Szebehely1981,Holman1999}.  However, this result is likely biased due to the large number of simulations we performed ($\sim$40\%) where $\mu \lesssim 0.2$ and Hill stability would be more applicable.  The largest values of the critical semimajor axis, $a_c$ in units of $a_{bin}$, occur for large binary eccentricity ($e_{bin}\sim 0.8$) and a more modest stellar mass ratio $\mu \sim 0.18$.  {Recently \cite{Lam2018} performed a similar study using machine learning through a deep neural network (DNN), where we find good agreement between the studies (typically within 5\%) when $\mu \gtrsim 0.05$ and much larger disagreement (up to $\sim$33\%) for smaller $\mu$.  They did not train for $\mu < 0.05$ and thus one should not use their DNN on such systems (D. Kipping, private communication).}

We apply 3 different methods to estimate the stability limit and compare to numerical simulations that take a wide range of values in the initial semimajor axis and eccentricity of the planet into account.  The derived stability limits for $a_c$ agree well (within $\sim$1\%) when considering either an ideal or more realistic architecture for the known Kepler CBPs.  The derived limits for $a_c$ can also be generalized to include eccentric planetary orbits by considering a proportional critical eccentricity, $e_c$.   

Our analysis also finds that {55\% of CBP systems from Kepler could host another equal-mass planet closer to $a_c$ on a coplanar orbit using numerical simulation and a planet packing framework \citep{Kratter2014}}.  We consider this to be a conservative estimate because smaller values of $a_c$ are possible if the interior planet is highly misaligned (or even retrograde) relative to the binary orbital plane \citep{Doolin2011,Li2016}.  As a result, we do \textit{not} find strong evidence for a pile-up near the stability limit for the Kepler CBP systems (see Table \ref{tab:pl_param}), especially considering the observing bias toward the discovery of small semimajor axis planets using conventional methods.  

However, we do find that most ($\sim$90\%) of the Kepler CBP host binary eccentricities are $<$0.25 and have similar stability limits ($a_c = 2.3 - 3.1 a_b$).  T{he dynamical spacing, $\beta_c$, is larger than 7 mutual Hill radii for the systems that could host an interior planet on coplanar orbit and indicates a need for more in-depth studies (Kepler-34, Kepler-413, Kepler-47, Kepler-453, \& Kepler-1647).}  Although the sample of confirmed Kepler CBPs is limited, observations from TESS are expected to substantially increase the sample, where we can then identify more robustly any trends within the CBP population in relation to the stability surface.

\acknowledgments
{We thank the anonymous referee for providing helpful comments that improved the overall quality and clarity of the manuscript.  The simulations presented here were performed using the OU Supercomputing Center for Education \& Research (OSCER) at the University of Oklahoma (OU).  S.S. would like to thank Zdzislaw Musielak and Alex Weiss for their continued support in the exoplanetary research.}

\bibliographystyle{apj}
\bibliography{refs}

\end{document}